\documentclass[reprint,aps,prl,amsmath,amssymb,showpacs]{revtex4-1}

\usepackage{graphicx}
\usepackage{dcolumn}
\usepackage{bm}
\usepackage{amsmath}
\usepackage{color}

\allowdisplaybreaks[2]

\newcommand{\ew}[1]{\langle #1 \rangle}				

\newcommand{\hide}[1]{}

\begin{document}

\title{
Inducing and detecting collective population inversions of M\"ossbauer nuclei
}

\author{K. P. Heeg}
\author{C. H. Keitel}
\author{J. Evers}
\affiliation{Max-Planck-Institut f\"{u}r Kernphysik, Saupfercheckweg 1, D-69117 Heidelberg, Germany}

\date{14.07.2016}

\begin{abstract}
Up to now, experiments involving M\"ossbauer nuclei driven by x-rays have been restricted to the low-excitation regime. Here, a setup is proposed which promises significant excitation, ideally exceeding full inversion of the nuclear ensemble, at x-ray light sources under construction. We further introduce a method to experimentally verify such inversions, in which population inversions manifest themselves in symmetry flips of suitably recorded spectra. It neither requires per-shot spectra of the incoming x-ray pulses, nor absolute measurements of the scattered light intensity. 
\end{abstract}

\maketitle

\allowdisplaybreaks[3]

Nuclear quantum optics by now is a very active research field, both theoretically~\cite{PhysRevLett.82.3593,Buervenich2006,Liao2012,Heeg2013b,tenBrinke2013,Liao2015,Gunst2015,Longo2016} and experimentally~\cite{Shvydko1996,Coussement2002,Roehlsberger2010,Roehlsberger2012,Heeg2013,Vagizov2014,Heeg2015,Heeg2015b,Haber2016}.
More related references can be found in~\cite{Adams2003,Roehlsberger2005,Adams2013,RESreview}.
Up to day, however, all experiments were restricted to the low-excitation regime, in which the number of excitations is negligible compared to the number of nuclei in the ensemble. This severely restricts the possible applications as compared to quantum optics operating at optical frequencies. For instance, efficient population transfer between nuclear states or significant manipulation of the nuclear level structure by strong control fields are currently unfeasible. In turn, non-linear light-matter interactions cannot be observed. 

A second problem arises from the fact that it is difficult to reliably detect a possible population inversion. In principle, the Rabi flopping goes along with a periodic modulation of the scattered light intensity. But in particular if the light source enables excitation close to inversion, but not multiple Rabi oscillations, or if SASE FEL pulses with resonant intensity varying significantly from pulse to pulse are used, an absolute determination of the scattered light intensity as function of the resonant incidence intensity is challenging.

Here, we propose a setup which solves both problems. First, we introduce a robust method to determine the population inversion using spectral interference, rather than via absolute intensity measurements. Second, the setup promises strong excitation of nuclear ensembles at projected x-ray sources, ideally exceeding full inversion. 

\begin{figure}[b]
 \centering
 \includegraphics[width=0.85\columnwidth]{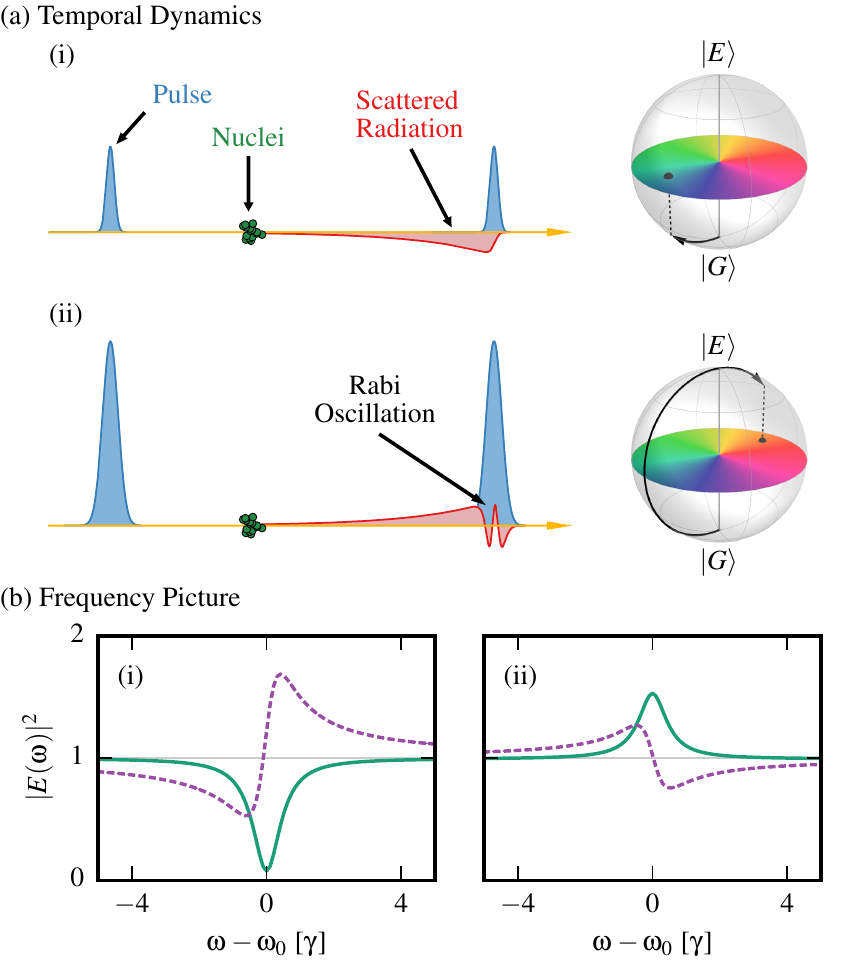}
 \caption{Method to detect population inversion. An incident pulse excites the nuclei, which subsequently emit scattered radiation. The observable spectra arise from the interference between incident and scattered radiation. As visualized in (i) for weak excitation and (ii) for excitation slightly beyond inversion, the induced dipole moment flips its sign whenever half a Rabi cycle is completed. These flips directly manifest themselves in the observed spectra. This way, population inversions can be detected in a direct and robust way.}
 \label{fig:toy}
\end{figure}

The main idea of the detection is illustrated in Fig.~\ref{fig:toy}. An incident x-ray pulse (blue) excites the nuclei, which subsequently emit the scattered radiation (red). A detector registers the incident pulse together with the scattered light, including the interference of the two contributions.
In the relevant case of a single nucleus excited by a short resonant $\delta$-like x-ray pulse $\mathcal{E}(t) = A_0 \delta(t)$, the electric field behind the scatterer can be written as~\cite{Agarwal1974}
\begin{align}
 {E(t)} = \mathcal{E}(t) + i \beta \, \ew{\hat{d}(t)}\;. \label{eq:toy_free_plus_scattered}
\end{align}

Here, the second part is the scattered light, where the dipole operator $\hat{d}$ with magnitude $d$ characterizes the nuclear excitation and $\beta$ is a constant depending on the sample density and size~\cite{Heidmann1985}. The short x-ray pulse drives the system into the state $\ew{\hat{d}(t)} = (i d/2)\sin{(\Phi)}$, where $\Phi = 2 d A_0$ denotes the pulse area. This state subsequently decays spontaneously with decay rate $\gamma$. Thus, 
\begin{align}
 {E(t)} &= A_0 \delta(t) -\frac{\beta d}{2} \sin{(\Phi)} \, \theta(t) e^{-\tfrac{\gamma}{2}t}\;,
\end{align}
where the step function $\theta(t) $ accounts for the excitation at $t=0$.
A Fourier transformation yields the coherent part of the observed light spectrum
\begin{align}
 |E(\omega)|^2
 &\sim \left| A_0 -  \frac{\tfrac{i}{2}\beta d \sin{(\Phi)}}{\omega - \omega_0 + i\tfrac{\gamma}{2}} \right|^2 \;. \label{eq:toy_spectrum}
\end{align}
We thus find that the incident pulse interferes with the scattered light, and the relative phase depends on the degree of excitation $\Phi$. For a weak pulse with area $\Phi \ll 1$, the system is only slightly perturbed out of the ground state, as illustrated via the Bloch sphere in Fig.~\ref{fig:toy}(a,i). As expected, the corresponding spectrum is a Lorentz absorption profile, shown in (b,i). But if the incident pulse is sufficiently strong to invert the population of the nuclei as in panel (a,ii), the factor $\sin(\Phi)$ flips its sign whenever half a Rabi cycle is completed. Accordingly, also the interference with the incident field and the resulting spectrum changes. This in turn leads to a conversion of the Lorentz absorption profile into a gain profile (b,ii). As a consequence, the population inversion clearly manifests itself in a qualitative change in the spectrum, rather than in a quantitative change in scattered light intensity.

We now turn to the setup to observe nuclear inversions. We base our considerations on x-ray cavities, which have proven to be a fruitful platform for nuclear quantum optics~\cite{Roehlsberger2010,Roehlsberger2012,Heeg2013,Heeg2013b,Heeg2015,Heeg2015b}. In particular it was already shown that it enables an enhancement of the interaction of x-rays with nuclei via cavity and coherence-based effects in a single setup~\cite{Heeg2015}. Also, the x-ray incidence angle onto the cavity controls a constant relative phase between the scattered and the unscattered contributions~\cite{Heeg2015b}, enabling further manipulation of the intensity-dependent spectra. As an example used in the following, the dashed lines in Fig.~\ref{fig:toy}(b) show the case of an additional $90^\circ$ phase shift, giving rise to Fano-like spectra~\cite{Fano1961,Ott2013}.

The state-of-the-art quantum optical description for x-ray cavities~\cite{Heeg2013b,Heeg2015c} is restricted to single excitations. Here, we extend this model to arbitrary excitations such that it encompasses population inversions. Introducing 
collective operators $\hat{J}_{\pm} = \sum_n \hat{S}_{\pm}^{(n)}$ as sum over all single particle raising/lowering operators, the Hamiltonian for the full system of $N$ nuclei after adiabatically eliminating the cavity mode is~\cite{Heeg2013b}
\begin{align}
 \hat{H} = \left(\xi a_\textrm{in}(t) \hat{J}_+ + \textrm{h.c.}\right) + \operatorname{Im}(\zeta) \hat{J}_+ \hat{J}_- \;. \label{eq:Hamiltonian}
\end{align}
Here, $a_\textrm{in}(t)$ is the amplitude of the driving field, $\zeta = 2 |g|^2/ 3 (\kappa + i\Delta_C)$ and $\xi = \zeta \sqrt{3\kappa_R} / g^*$, where $\kappa$ is the cavity decay rate, $\Delta_C$ the cavity detuning, and $\kappa_R$ and $g$ the coupling coefficients of the cavity mode to the external field and the nuclei, respectively. The two parts in Eq.~(\ref{eq:Hamiltonian}) account for driving field and cooperative Lamb shift. In a master equation approach, spontaneous emission (SE) and superradiant decay (SR) are governed by the Lindblad terms
\begin{align}
 \mathcal{L}_\textrm{SE} &= -\frac{\gamma}{2} \sum_n \left( \hat{S}_+^{(n)} \hat{S}_-^{(n)} \hat{\rho} -  \hat{S}_-^{(n)} \hat{\rho} \hat{S}_+^{(n)}  + \textrm{H. c.}\right)  \label{eq:Lindblad_SE} \;, \\
 \mathcal{L}_\textrm{SR} &= -\operatorname{Re}(\zeta)\left( \hat{J}_+ \hat{J}_- \hat{\rho}  -  \hat{J}_- \hat{\rho} \hat{J}_+ + \textrm{H. c.}\right) \;, \label{eq:Lindblad_SR}
\end{align}
where $\hat{\rho}$ is the ensemble density matrix.

In~\cite{Heeg2013b,Heeg2015c}, these equations were solved in linear response with at most one excitation in the system. Extending to arbitrary excitations, we are faced with the problem that the system's Hilbert space dimension is $2^N$, rendering direct calculations impossible. To overcome this problem, we make use of the fact that the observed signal is dominated by the coherent part of the scattered light, which is highly collimated in forward direction and scales as $N^2$~\cite{Roehlsberger2005}. This part arises from the collective superradiant nuclear dynamics on a time scale fast compared to that of the incoherent single particle spontaneous decay. We can thus approximately neglect the single particle spontaneous decay $\mathcal{L}_\textrm{SE}$. With this approximation, we can map our system onto a Dicke model~\cite{Dicke1954}, in which only the $N+1$ symmetric states are mutually coupled. This reduces the Hilbert space dimension from $2^N$ to $N+1$ states. Using standard input-output relations, we can then evaluate the time-dependent field reflected by the cavity as
$\hat{a}_\textrm{out}(t) = [2\kappa_R/(\kappa + i\Delta_C) -1 ] \, a_\textrm{in}(t) - i \xi\, \hat{J}_-(t)$ for arbitrary input fields $a_\textrm{in}(t)$. With $\hat{a}_\textrm{out}(\omega)$ the Fourier transform of $\hat{a}_\textrm{out}(t)$, the spectrum of the coherently scattered light is $I_\textrm{coh} = |\ew{\hat{a}_\textrm{out}(\omega)}|^2$~\cite{Eberly1992}.
Our numerical calculations typically are performed with $N\sim100$, which is significantly lower than the typical number of involved nuclei. However, we found that this number of atoms captures the many-body character well, and allows to extrapolate results for larger $N$ via a suitable scaling of the coupling rate $g$ (see Supplemental Material).
Finally, we note that the input-output relation has the same structure as our toy model Eq.~(\ref{eq:toy_free_plus_scattered}): The empty cavity response replaces the incident field part, and the collective dipole operator of the ensemble represents the scattered light.

\begin{figure}[t]
 \centering
 \includegraphics[width=0.9\columnwidth]{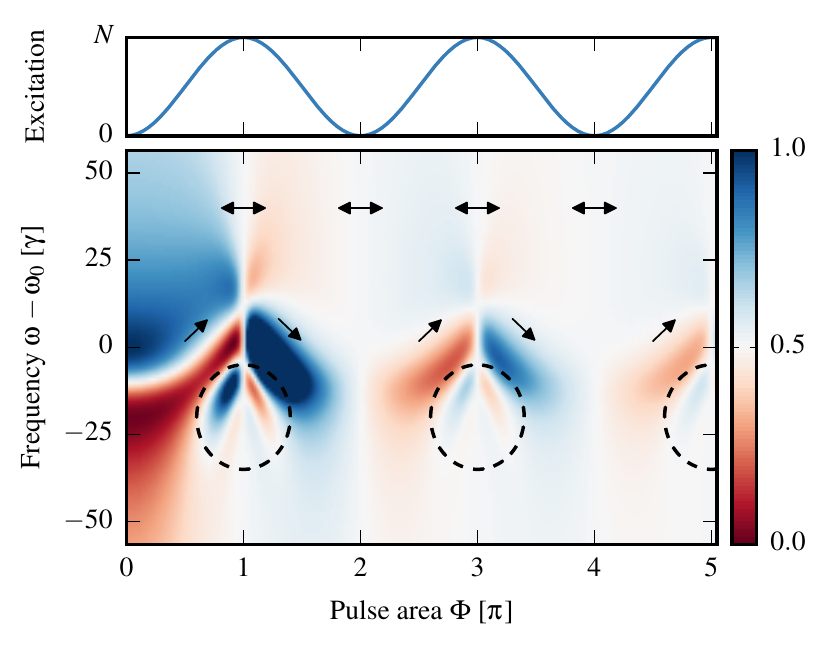}
 \caption{Normalized x-ray spectra $I_\textrm{coh} / I_\textrm{in}$ after excitation with a Fourier-limited Gaussian pulse. With increasing x-ray intensity, the symmetry of the spectrum flips repeatedly (double headed arrows), indicating Rabi oscillations of the nuclei. The nuclear excitation directly after the pulse is shown in the upper panel.}
 \label{fig:gauss}
\end{figure}

Next, we present our numerical results. To connect with previous work, we first use a layer system  Pt(2.6~nm)/C(7.9~nm)/$^{57}$Fe(1.5~nm)/C(9.3~nm)/Pt, which is the cavity studied in~\cite{Heeg2013b} at low excitation. Below, we will optimize this cavity structure systematically. The resonant nucleus is the isotope $^{57}$Fe with transition energy $\omega_0 = 14.4$~keV and life time $1/\gamma = 141$~ns. Based on the absorption length in the guiding layer of the cavity we estimate a coherence volume with $N\approx 10^{11}$ nuclei involved in the collective dynamics~(see Supplementary Material). Parameters $\kappa$, $\kappa_R$ and $g\sqrt{N}$ are chosen as in~\cite{Heeg2013b}, further we set $\Delta_C =\kappa$.

To verify our new detection scheme, we first calculated spectra for Gaussian input pulses $a_\textrm{in}(t) \sim \exp{ (-t^2 / 2\sigma_t^2)}$ with $\sigma_t = 100$~fs. Results normalized to the spectrum of the input pulses $I_\textrm{in} = |a_\textrm{in}(\omega)|^2$ are shown in Fig.~\ref{fig:gauss}. The pulse amplitude is characterized via its area $\Phi = 2 |\xi| \int a_\textrm{in}(t) dt$. As predicted, the spectrum changes its symmetry whenever $\Phi$ exceeds a multiple of $\pi$, i.e., whenever half a Rabi cycle is completed (double headed arrows). This can be best seen from the asymptotic behavior at large frequencies. However, compared to our simple toy model, additional features appear in the center of the spectrum. First, the spectra are shifted relative to the nuclear resonance, which is due to the cooperative Lamb shift (single headed arrows). More interestingly, there are additional minima and maxima in the spectrum if the number of excited nuclei exceeds $N/2$ (dashed circles). We found that these arise due to the modified temporal decay of the nuclear ensemble characteristic of superradiance~\cite{Gross1982}. If more than half of the nuclei are excited, they do not decay exponentially like in the weak excitation case. Rather, the well-known intensity burst at a finite time after excitation occurs. This modified temporal decay in turn induces the additional structures in the spectrum. This feature is highly favorable, since it not only signifies collective dynamics, but also enables a clear determination of strong excitation of the nuclear ensemble even if the light source does not enable full inversion of the ensemble.
Finally, we note that the spectra shown in Fig.~\ref{fig:gauss} become flat for $\Phi = n\cdot\pi$. The reason is that the off-diagonal elements of the density matrix, which give rise to coherent emission, vanish if the system is either in the ground or in the fully excited state.

\begin{figure}[t]
 \centering
 \includegraphics[width=0.9\columnwidth]{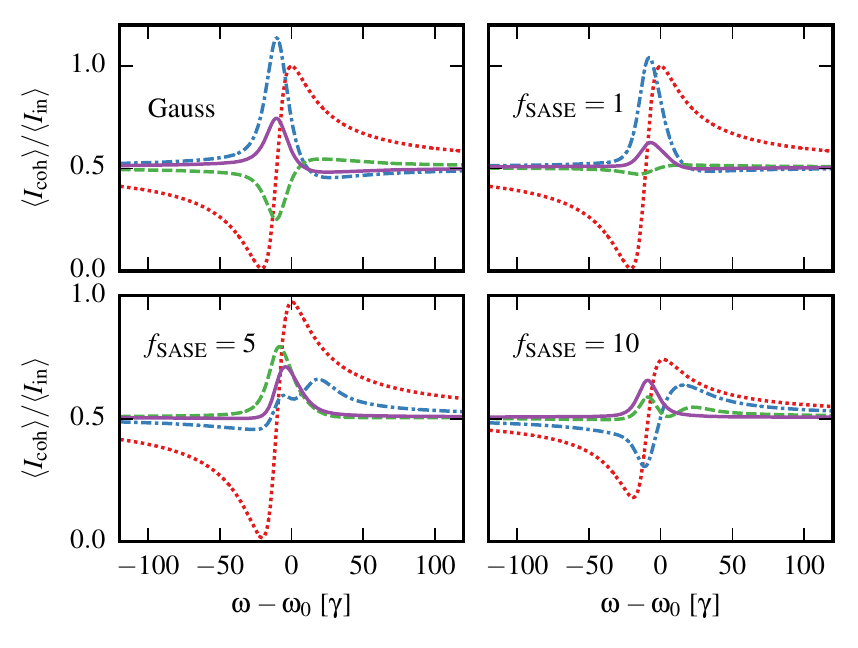}
 \caption{Normalized spectra, averaged over $100$ SASE pulses with $f_\textrm{SASE} = \sigma_\omega \sigma_t$. [Dotted red/dash-dotted blue/dashed green/solid purple] curves correspond to maximum pulse areas $\Phi_\textrm{max}=$ [$\tfrac{1}{50}$/$\tfrac{3}{2}$/$\tfrac{5}{2}$/$\tfrac{7}{2}$]$\times \pi$. Even for large $f_\textrm{SASE}$, signatures of symmetry flipping, i.e.~Rabi oscillations are still visible.}
 \label{fig:sase}
\end{figure}

After having established our detection method for Gaussian input pulses, which now turn to SASE x-ray pulses. These are routinely available in current x-ray facilities, but are neither Fourier-limited nor have a Gaussian shape. We model the stochastic SASE pulses employing the partial coherence method~\cite{Pfeifer2010}, assuming Gaussian mean pulse shapes in the time and frequency domains, with widths $\sigma_t = 100$~fs and $\sigma_\omega = f_\textrm{SASE}/\sigma_t$. A larger number $f_\textrm{SASE}$ therefore results in noisier pulses.

Results for SASE pulses are shown in Fig.~\ref{fig:sase}. In experiments, it is challenging to record spectra before and after interaction with the target on a per-shot basis~\cite{Yudovich2014}. For this reason, we obtained our results by averaging the results for $100$ different pulses with the same {\it total} intensity before the interaction. This intensity is typically experimentally accessible for each pulse, and we quantify it in terms of an effective pulse area $\Phi_\textrm{max}$, which is the area of a Fourier-limited Gaussian pulse with the same total intensity. Further, the results are normalized to the average input pulse spectra.

From Fig.~\ref{fig:sase} we find that the characteristic change in the spectrum symmetry can be observed even for large $f_\textrm{SASE}$. As expected, the spectral response fluctuates due to the stochastic nature of the SASE pulses, and averaging over fluctuating shots results in a decreased visibility. As a consequence, the intensities required to clearly observe nuclear inversions increase, see Fig.~\ref{fig:sase}. Note that the required mean intensity reduces if the resonant intensities before and after the interaction could be measured on a per-shot basis, since then  pulses with similar resonant intensity could be grouped in the averaging.

Finally, we analyze the absolute pulse intensities required in order to induce a full population inversion, and compare it to intensities of existing and projected x-ray sources. The condition for full inversion in terms of the pulse area is $\Phi \ge \pi$. For a Fourier-limited pulse of duration $\sigma_t$, the integral in the pulse area $\int a_\textrm{in} dt \sim \sqrt{N_\textrm{Ph} \sigma_t}$. As expected, longer pulses are favorable, since they correspond to a more narrow pulse spectrum. However, narrowing of the spectrum is limited due to the angular beam divergence, since it leads to a detuning with respect to the cavity mode resonance, effectively increasing the spectral width again. Hence, the condition to observe a population inversion can be expressed as a minimal number of required photons $N_\textrm{Ph}$, which depends on beam divergence, pulse duration and the cavity structure (see Supplemental Material).

\begin{figure}[t]
 \centering
 \includegraphics[width=0.9\columnwidth]{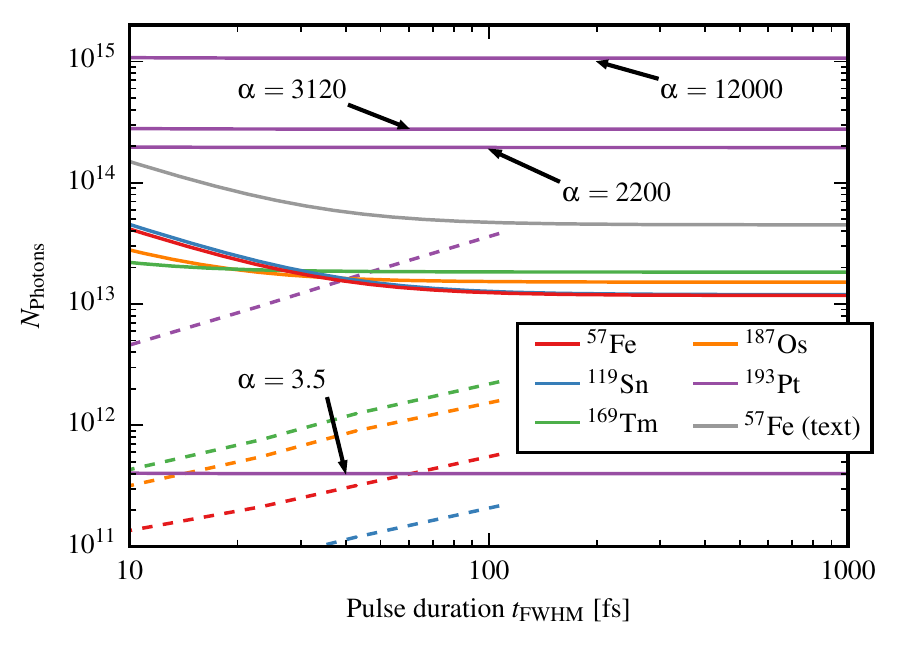}
 \caption{Comparison of population inversion requirements for different resonant nuclei in optimized cavities as well as the cavity employed above. Beam parameters of the European XFEL at the respective transition energies are shown as dashed lines. Data for $^{193}$Pt is shown for the different internal conversion coefficients $\alpha$ found in the literature.}
 \label{fig:nphmin}
\end{figure}

In addition to $^{57}$Fe, we consider several other isotopes with low-lying transition energies and calculate the number of required photons for different cavity layouts with the aim to relax the requirements for the x-ray sources. This optimization is challenging, since several parameters and a number of effects need to be considered. For example, the thickness of the topmost layer $d_{\textrm{top}}$, which is acting as mirror, influences the quality factor $Q$ of the cavity mode. A high $Q$ leads to an increased effective coupling strength, but also corresponds to narrow spectral widths of the cavity mode, which effectively reduces the angular acceptance for the incoming x-ray field. This in turn restricts the possibility to focus the x-ray beam, and therefore enlarges the number of involved nuclei, rendering full inversion more challenging. Apart from that, $d_{\textrm{top}}$ controls the visibility of the interference between the scattered and the incoming light, which is our main observable. In our numerical study, the layer thicknesses and type of mirror materials were varied and the focal spot of the beam was chosen such that no other cavity mode is driven due to the beam divergence (see Supplementary Material).
The result of our optimization is shown in Fig.~\ref{fig:nphmin}. As visible for the case of $^{57}$Fe, optimizing the cavity structure provides a handle to lower the requirements for a population inversion. Since the degree of nuclear excitation scales with the pulse area as $\sin^2(\Phi /2)$, our model predicts excitation fractions of order $\sim 10\%$ for the XFEL parameters and the optimized structures.
At the upcoming European XFEL, a full inversion could already be observed for the isotope $^{193}$Pt, depending on the internal inversion coefficient $\alpha$ for which different values were reported in the literature~(see Supplementary Material). At the $^{193}$Pt resonance energy $1.642$~keV, pulses with $4\times 10^{13}$ photons and duration $107$~fs FWHM are predicted~\cite{Schneidmiller2011}, which according to Fig.~\ref{fig:nphmin} should be sufficient to achieve inversion for $\alpha = 3.5$. Excitation fractions of $\sim 40\%$ and $30\%$ are predicted for $\alpha = 2200$ and $\alpha = 3120$, respectively. Similar machine parameters are achieved at LCLS~\cite{Emma2010} or SACLA~\cite{Ishikawa2012}. We expect that the required coherence quality of x-ray pulses can likely be reached by utilizing self-seeding schemes~\cite{Altarelli2007,Geloni2011,Amann2012} in combination with high-resolution monochromators~\cite{Toellner2011}. Note that our results also show that the cavity approach enables one to enhance the excitation of nuclear ensembles by orders of magnitude compared to the free space case~\cite{Junker2012}.

An interesting alternative could be future XFELO sources, which will generate near Fourier-limited pulses by design~\cite{Kim2008,Lindberg2011,Adams2015}. These feature a lower number of photons per pulse ($\sim 10^9$), such that focusing is mandatory in order to reduce the size of the nuclear ensemble below the number of resonant photons in the x-ray pulse. The XFELO has the unique advantage that subsequent pulses with a MHz repetition rate are mutually phase coherent, such that the full inversion could be achieved using a sequence of multiple XFELO pulses if the ratio of pulse separation and nuclear life time is favorable. A particularly interesting candidate in this respect and for applications in precision metrology and spectroscopy is ${}^{45}$Sc with a very long lifetime of $470$ms, corresponding to a feV linewidth.  Estimating that the excitation is given by the combined effect of pulses throughout the lifetime, the XFELO promises excitation fractions on the percent level, which is about six orders of magnitude higher than for XFEL parameters.

In summary, we presented a robust method to reliably detect the inversion or Rabi-oscillations of nuclear ensembles excited by x-ray pulses. It neither requires per-shot spectra of the x-ray pulses, nor an absolute measurement of the scattered light intensity. We further proposed a setup which promises significant population inversion of nuclear ensembles at projected x-ray sources. These results open new avenues for x-ray quantum optics, involving significant population transfer between nuclear states, significant manipulation of the nuclear level structure by strong control fields, and non-linear light-matter interactions.

\bibliographystyle{myprsty}
\bibliography{nonlinear}

\begin{thebibliography}{10}

\bibitem{PhysRevLett.82.3593}
O. Kocharovskaya, R. Kolesov, and Y. Rostovtsev, Phys. Rev. Lett. {\bf 82},
  3593  (1999).

\bibitem{Buervenich2006}
T.~J. B\"urvenich, J. Evers, and C.~H. Keitel, Phys. Rev. Lett. {\bf 96},
  142501  (2006).

\bibitem{Liao2012}
W.-T. Liao, A. P\'alffy, and C.~H. Keitel, Phys. Rev. Lett. {\bf 109},  197403
  (2012).

\bibitem{Heeg2013b}
K.~P. Heeg and J. Evers, Phys. Rev. A {\bf 88},  043828  (2013).

\bibitem{tenBrinke2013}
N. ten Brinke, R. Sch\"utzhold, and D. Habs, Phys. Rev. A {\bf 87},  053814
  (2013).

\bibitem{Liao2015}
W.-T. Liao and S. Ahrens, Nature Photonics {\bf 9},  169  (2015).

\bibitem{Gunst2015}
J. {Gunst}, C.~H. {Keitel}, and A. {P{\'a}lffy}, Scientific Reports {\bf 6},
  25136  (2016).

\bibitem{Longo2016}
P. Longo, C.~H. Keitel, and J. Evers, Scientific Reports {\bf 6},  23628
  (2016).

\bibitem{Shvydko1996}
Y.~V. Shvyd'ko, T. Hertrich, U. van B\"urck, E. Gerdau, O. Leupold, J. Metge,
  H.~D. R\"uter, S. Schwendy, G.~V. Smirnov, W. Potzel, and P. Schindelmann,
  Phys. Rev. Lett. {\bf 77},  3232  (1996).

\bibitem{Coussement2002}
R. Coussement, Y. Rostovtsev, J. Odeurs, G. Neyens, H. Muramatsu, S. Gheysen,
  R. Callens, K. Vyvey, G. Kozyreff, P. Mandel, R. Shakhmuratov, and O.
  Kocharovskaya, Phys. Rev. Lett. {\bf 89},  107601  (2002).

\bibitem{Roehlsberger2010}
R. R\"ohlsberger, K. Schlage, B. Sahoo, S. Couet, and R. R\"uffer, Science {\bf
  328},  1248  (2010).

\bibitem{Roehlsberger2012}
R. R\"ohlsberger, H.-C. Wille, K. Schlage, and B. Sahoo, Nature {\bf 482},  199
   (2012).

\bibitem{Heeg2013}
K.~P. Heeg, H.-C. Wille, K. Schlage, T. Guryeva, D. Schumacher, I. Uschmann,
  K.~S. Schulze, B. Marx, T. K\"ampfer, G.~G. Paulus, R. R\"ohlsberger, and J.
  Evers, Phys. Rev. Lett. {\bf 111},  073601  (2013).

\bibitem{Vagizov2014}
F. Vagizov, V. Antonov, Y.~V. Radeonychev, R.~N. Shakhmuratov, and O.
  Kocharovskaya, Nature {\bf 508},  80  (2014).

\bibitem{Heeg2015}
K.~P. Heeg, J. Haber, D. Schumacher, L. Bocklage, H.-C. Wille, K.~S. Schulze,
  R. Loetzsch, I. Uschmann, G.~G. Paulus, R. R\"uffer, R. R\"ohlsberger, and J.
  Evers, Phys. Rev. Lett. {\bf 114},  203601  (2015).

\bibitem{Heeg2015b}
K.~P. Heeg, C. Ott, D. Schumacher, H.-C. Wille, R. R\"ohlsberger, T. Pfeifer,
  and J. Evers, Phys. Rev. Lett. {\bf 114},  207401  (2015).

\bibitem{Haber2016}
J. Haber, K.~S. Schulze, K. Schlage, R. Loetzsch, L. Bocklage, T. Gurieva, H.
  Bernhardt, H.-C. Wille, R. R\"uffer, I. Uschmann, G.~G. Paulus, and R.
  R\"ohlsberger, Nat Photon {\bf 10},  445  (2016).

\bibitem{Adams2003}
{\em Nonlinear Optics, Quantum Optics, and Ultrafast Phenomena with X-Rays:
  Physics with X-Ray Free-Electron Lasers}, edited by B.~W. Adams (Springer,
  Heidelberg, 2003).

\bibitem{Roehlsberger2005}
R. R\"ohlsberger, {\em Nuclear Condensed Matter Physics with Synchrotron
  Radiation}, Vol.~208 of {\em Springer Tracts in Modern Physics} (Springer,
  Berlin Heidelberg, 2005).

\bibitem{Adams2013}
B.~W. Adams, C. Buth, S.~M. Cavaletto, J. Evers, Z. Harman, C.~H. Keitel, A.
  P\'alffy, A. Picon, R. R\"ohlsberger, Y. Rostovtsev, and K. Tamasaku, Journal
  of Modern Optics {\bf 60},  2  (2013).

\bibitem{RESreview}
R. R\"ohlsberger, J. Evers, and S. Shwartz,  in {\em Synchrotron Light Sources
  and Free-Electron Lasers}, edited by E. Jaeschke, S. Khan, J.~R. Schneider,
  and J.~B. Hastings (Springer, Berlin, 2015), pp.\ 1--28.

\bibitem{Agarwal1974}
G. Agarwal, {\em Quantum Optics: Quantum Statistical Theories of Spontaneous
  Emission and Their Relation to Other Approaches}, {\em Springer tracts in
  modern physics} (Springer, Berlin, Heidelberg, 1974).

\bibitem{Heidmann1985}
A. Heidmann and S. Reynaud, Journal de Physique {\bf 46},  1937  (1985).

\bibitem{Fano1961}
U. Fano, Phys. Rev. {\bf 124},  1866  (1961).

\bibitem{Ott2013}
C. Ott, A. Kaldun, P. Raith, K. Meyer, M. Laux, J. Evers, C.~H. Keitel, C.~H.
  Greene, and T. Pfeifer, Science {\bf 340},  716  (2013).

\bibitem{Heeg2015c}
K.~P. Heeg and J. Evers, Phys. Rev. A {\bf 91},  063803  (2015).

\bibitem{Dicke1954}
R.~H. Dicke, Phys. Rev. {\bf 93},  99  (1954).

\bibitem{Eberly1992}
J.~H. Eberly and M.~V. Fedorov, Phys. Rev. A {\bf 45},  4706  (1992).

\bibitem{Gross1982}
M. Gross and S. Haroche, Physics Reports {\bf 93},  301  (1982).

\bibitem{Pfeifer2010}
T. Pfeifer, Y. Jiang, S. D\"{u}sterer, R. Moshammer, and J. Ullrich, Opt. Lett.
  {\bf 35},  3441  (2010).

\bibitem{Yudovich2014}
S. Yudovich and S. Shwartz, Phys. Rev. A {\bf 90},  033805  (2014).

\bibitem{Schneidmiller2011}
E.~A. Schneidmiller and M.~V. Yurkov, {P}hoton beam properties at the
  {E}uropean {XFEL}, 2011.

\bibitem{Emma2010}
P. Emma {\it et~al.}, Nature Photonics {\bf 4},  641  (2010).

\bibitem{Ishikawa2012}
T. Ishikawa {\it et~al.}, Nature Photonics {\bf 6},  540  (2012).

\bibitem{Altarelli2007}
{\em The European X-Ray Free-Electron Laser: Technical design report},
  No.~2006-097 in {\em DESY}, edited by M. Altarelli {\it et~al.} (DESY XFEL
  Project Group, Hamburg, 2007).

\bibitem{Geloni2011}
G. Geloni, V. Kocharyan, and E. Saldin, ArXiv e-prints  (2011).

\bibitem{Amann2012}
J. Amann {\it et~al.}, Nature Photonics {\bf 6},  693  (2012).

\bibitem{Toellner2011}
T.~S. Toellner, A. Alatas, and A.~H. Said, Journal of Synchrotron Radiation
  {\bf 18},  605  (2011).

\bibitem{Junker2012}
A. Junker, A. P\'alffy, and C.~H. Keitel, New Journal of Physics {\bf 14},
  085025  (2012).

\bibitem{Kim2008}
K.-J. Kim, Y. Shvyd'ko, and S. Reiche, Phys. Rev. Lett. {\bf 100},  244802
  (2008).

\bibitem{Lindberg2011}
R.~R. Lindberg, K.-J. Kim, Y. Shvyd'ko, and W.~M. Fawley, Phys. Rev. ST Accel.
  Beams {\bf 14},  010701  (2011).

\bibitem{Adams2015}
B.~W. Adams and K.-J. Kim, Phys. Rev. ST Accel. Beams {\bf 18},  030711
  (2015).

\bibitem{Hannon1999}
J. Hannon and G. Trammell, Hyperfine Interactions {\bf 123-124},  127  (1999).

\bibitem{Scully1997}
M. Scully and S. Zubairy, {\em Quantum Optics} (Cambridge University Press,
  Cambridge, 1997).

\bibitem{Seto2011}
M. Seto,  in {\em Handbook of Nuclear Chemistry}, edited by A. V{\'e}rtes, S.
  Nagy, Z. Klencs{\'a}r, R.~G. Lovas, and F. R{\"o}sch (Springer, Boston, MA,
  2011), Chap.~M{\"o}ssbauer Excitation by Synchrotron Radiation, pp.\
  1447--1460.

\bibitem{Parratt1954}
L.~G. Parratt, Phys. Rev. {\bf 95},  359  (1954).

\bibitem{Alder1972}
K. Alder, U. Raff, and G. Baur, Helv. Phys. Acta {\bf 45},  771  (1972).

\bibitem{NNDC}
{N}ational {N}uclear {D}ata {C}enter, http://www.nndc.bnl.gov/, accessed:
  2016-07-06.

\bibitem{Kibedi2008}
T. Kib\'edi, T. Burrows, M. Trzhaskovskaya, P. Davidson, and C. {Nestor Jr.},
  Nuclear Instruments and Methods in Physics Research Section A: Accelerators,
  Spectrometers, Detectors and Associated Equipment {\bf 589},  202  (2008).

\end{thebibliography}
 \clearpage

\section{Supplemental Material}

\subsection{Rescaling of the coupling constant $g$}
Even after mapping our system to a Dicke model as explained in the main text, the size of the differential equation system prohibits a numerical simulation of all nuclei driven by a focused x-ray beam. We circumvent this obstacle by reducing the number of particles, but at the same time adjust the coupling rates, such that the results remain unchanged. In the linear limit, the collective coupling rate $g\sqrt{N}$ characterizes all observables including the spectra. Hence, this quantity must be kept invariant. We gradually increase the number of nuclei, while accordingly decreasing the value of $g$. Starting from $N\approx 5$, the many-body effect of the superradiant decay burst becomes visible and its structure in the spectrum quickly converges with respect to $N$. In the main text, we show results obtained for $N=100$, where we can expect that the many-body characteristics are fully captured.

\subsection{Number of interacting nuclei}
The cavity geometry for the resonant nuclei, which we analyze in the main text, facilitates the occurrence of cooperative effects such as superradiance and the collective Lamb shift, as experimentally shown in~\cite{Roehlsberger2010}. However, somehow interestingly, it is not directly known how many nuclei actually take part in this collective behavior. In the different theoretical and numerical approaches it is usually the density which enters the calculations~\cite{Hannon1999}, but not the corresponding volume which would allow to deduce the number of nuclei $N$. In the quantum optical model~\cite{Heeg2013b} the problem is similar, as for results in the linear regime only the combined quantity $g\sqrt{N}$, where $g$ is the coupling constant between the cavity mode and the nuclear transition, enters the observables. Hence, from comparison with experimental data, only $g\sqrt{N}$, but not the respective values of $g$ and $N$ can be extracted.

In a first approach we estimate the volume in which the coherent dynamics between the different nuclei can occur. For a large atomic separation, the damping due to absorption in the layer materials of the cavity gives rise to dephasing between the dipole moments and the collective behavior vanishes. For the cavity studied in the main text, the resonant layer consists of iron, for which the absorption length for x-rays at $14.4$~keV is $d\approx 20~\mu$m~\cite{Hannon1999}. Alternatively, the length can be estimated via the maximum distance a photon can travel during the cavity mode lifetime $1/\kappa$. Again, this yields several $10~\mu$m. Thus, for the coherence volume we assume a disk with diameter $d$ and thickness given by the size of the iron layer. With a number density of $n\approx 83~$nm$^{-3}$ we find the number of collectively interacting nuclei $N\sim 10^{11}$.

A second method to determine the number of nuclei consists in estimating the coupling constant $g = d\sqrt{\omega / 2 \epsilon_0 V}$, where $d$ is the dipole moment of the nuclear transition and $V$ is the mode volume. The dipole moment can be found from comparing the radiative part of spontaneous decay with the Weisskopf formula for spontaneous emission~\cite{Scully1997}. For the mode volume we assume a disk with diameter $d$ for the same reasons as before, however, its thickness is now given by the full extent of the cavity, i.e.~the thickness of the guiding layer. With this approach we arrive again at $N\sim 10^{11}$.

A method to determine $N$ independent of our quantum optical model is enabled by the work of~\cite{Longo2016}, where superradiance and its dependence on the sample geometry is studied in detail. Interpreting the thin layer of resonant nuclei in the cavity as a quasi 2-dimensional structure, the number $N$, required to obtain the experimentally observed magnitude of cooperative effects, is well comparable to our previous estimations.

\subsection{Required x-ray pulse intensity}
In order to observe the symmetry flipping of the spectrum, at least half a Rabi cycle must be completed. The properties of the x-ray pulse, inducing such a Rabi oscillation, are estimated below. We restrict the discussion to Fourier-limited pulses, for which the pulse shape $a_\textrm{in}(t)$ is a real quantity. In this case, the condition for symmetry flipping can be written in terms of the pulse area
\begin{align}
 \Phi = 2 |\xi| \int a_\textrm{in}(t) dt \overset{!}{\ge} \pi \;.
\end{align}
Next, we estimate the value of the integral. The number of photons in the pulse is given by $\int |a_\textrm{in}(t)|^2 dt = N_\textrm{Ph}$. Assuming a pulse with Gaussian envelope and width $\sigma_t = t_\textrm{FWHM}/2.35$, we obtain  $\int a_\textrm{in} dt = \sqrt{2 \sqrt{\pi} N_\textrm{Ph} \sigma_t}$. From this it is already visible, that not only the pulse intensity, i.e.~the number of photons is of importance, but also a long pulse duration is desirable. This can be understood in the following qualitative picture. A large duration $\sigma_t$ corresponds to a pulse with a higher compression in frequency domain, and consequently, the photon energies are closer to the nuclear resonance frequency on average, thus enhancing the interaction probability. With the considerations so far, we arrive at the condition $8 N_\textrm{Ph} |\xi|^2 \sigma_t \pi^{-3/2} \ge 1$.

Additionally, the spot size and the beam divergence have to be considered in our estimation. Both are linked, since focusing of the beam results in a larger divergence. The phase space, i.e.~the product of beam diameter $d_B$ and the angle of divergence $\theta_B$, is constant and typical values can be found in~\cite{Schneidmiller2011}. Let us first discuss the effect of the spot size. The beam impinges under grazing incidence with angle $\theta_0$, such that a guided mode of the cavity is driven. The volume in which nuclei are excited is therefore $V_\textrm{exc} \sim d_B^2 \, d / \sin{(\theta_0)}$, where $d$ is the thickness of the resonant layer. This volume can differ from the coherence volume estimated above. To properly include the number of actually excited nuclei, the condition derived above has to be extended by an additional factor $V_\textrm{coh}/V_\textrm{exc}$. To relax the condition on the light sources, the excitation volume and therefore the number of excited atoms should be as small as possible. Decreasing the beam spot size with suitable optics in an experiment is beneficial, however this comes at the drawback of a larger beam divergence. If the beam divergence is too large, the incidence angle $\theta_0$ is effectively modified and higher order guided modes of the cavity can be driven. This limits the maximum divergence to $\theta_{B,\textrm{max}} \approx |\theta_1 - \theta_0|$, where $\theta_1$ is the resonance angle of the next closest cavity mode. Further, the spectral width of the x-ray pulse is effectively broadened, since the angular detuning from resonance angle $\theta_0$ due to the beam divergence results in a detuning from the cavity mode resonance. The effective pulse width becomes $\sigma_{\omega,\textrm{eff}} = \sqrt{\sigma_\omega^2 + (\omega_0 \theta_0 \theta_B)^2}$ with $\sigma_\omega = 1/\sigma_t$. Consequently, the pulse duration $\sigma_t$ from the condition above has to be replaced by the effective duration $\sigma_{t,\textrm{eff}} = 1/\sigma_{\omega,\textrm{eff}}$. For very short pulses, the Fourier limit determines the spectral width, while at longer durations the beam divergence becomes the dominant effect. Especially for the case of resonant $^{57}$Fe nuclei, these limiting cases can be observed in Fig.~4 in the main text.

\subsection{Numerical cavity optimization}
\begin{table}[t]
\caption{Optimization of the cavity structure for the implementation of full inversion. $E_0$ is the resonance energy of the isotope in keV, $M$ the mirror material, $d_{\textrm{top}}$ and $d_{\textrm{cen}}$ the thicknesses of the mirror and center layer in nm, $N_{\textrm{exc}}$ the number of nuclei in the excitation volume, $d_B$ and $\theta_B$ the beam diameter and divergence in $\mu$m and mrad, respectively, and $N_{\textrm{Ph}}$ is the required number of photons at a pulse length $t_\textrm{FWHM} = 100$~fs. The resonant isotope layer thickness is  1~nm.\label{tab:optimization}}
\begin{ruledtabular}
\begin{tabular}{c c c c c c c c c }
Isotope     & $E_0$ & M & $d_{\textrm{top}}$ & $d_{\textrm{cen}}$ & $N_{\textrm{exc}}$ & $d_B$ & $\theta_B$ & $N_{\textrm{Ph}}$   \\
${}^{57}$Fe  & 14.4  &  Pt      & 6     & 20      & 6$\times 10^{7}$  & 45     & 1.1     & 1$\times 10^{13}$      \\
${}^{193}$Pt & 1.64  &  Pd      & 2.5   & 8       & 7$\times 10^{5}$  & 19     & 21.2    & 4$\times 10^{11}$      \\
${}^{119}$Sn & 23.9  &  Pt      & 4     & 15      & 3$\times 10^{7}$  & 39     & 0.9     & 1$\times 10^{13}$      \\
${}^{169}$Tm & 8.41  &  Pt      & 4     & 8       & 2$\times 10^{6}$  & 19     & 4.1     & 2$\times 10^{13}$      \\
${}^{187}$Os & 9.75  &  Pd      & 3     & 14      & 1$\times 10^{7}$  & 32     & 2.0     & 2$\times 10^{13}$      
\end{tabular}
\end{ruledtabular}
\end{table}
To optimize the cavity structure, we used a numerical approach in which the required number of photons at fixed pulse duration $t_\textrm{FWHM} = 100$~fs was calculated. This optimization was performed for different resonant isotopes with relatively low-lying transition energies, for which the material parameters were taken from~\cite{Roehlsberger2005,Seto2011,Junker2012}. The thickness of the resonant layer was set to 1~nm, the guiding layer to carbon, and as mirror material both Pt and Pd were considered. For each configuration the beam spot size was minimized by setting the beam divergence to $\theta_{B,\textrm{max}}$. To determine the parameters $\kappa, \kappa_R, \theta_0$ and $g$ of our theoretical model, we fitted the linear limit of $I_\textrm{coh}$ (see~\cite{Heeg2013b}), which depends on the incidence angle and the frequency, to numerical data obtained with the well-established formalism of~\cite{Parratt1954}. The respective best cavity layouts are listed in Tab.~\ref{tab:optimization} and the photon requirements are shown in more detail in Fig.~4 in the main text. A very promising nucleus is $^{193}$Pt with its transition at energy $\omega_0 = 1.642$~keV, where the required photon number could be reached in current XFEL facilities. The possibility to observe a full population inversion, though, depends on the internal conversion coefficient $\alpha$, for which the different values $3.5$~\cite{Alder1972}, $2200$~\cite{NNDC}, $3120$~\cite{Kibedi2008} and $12000$~\cite{Seto2011} are reported in the literature. Since the nuclear scattering length scales with $(1+\alpha)^{-1}$~\cite{Hannon1999}, this affects the estimation for the required intensity in the same manner. In Tab.~\ref{tab:optimization} the calculation was performed with $\alpha=3.5$, Fig.~4 in the main text shows data for all four coefficients.

\end{document}